\documentclass[showpacs,preprintnumbers,superscriptaddress,amsmath,amssymb]{revtex4}
\begin{document}
%
\bibliographystyle{try}

\title{Search for Medium Modifications of the $\rho$ Meson}

\newcommand*{\SCAROLINA}{University of South Carolina, Columbia, South Carolina 29208}
\affiliation{\SCAROLINA}
\newcommand*{\FIU}{Florida International University, Miami, Florida 33199}
\affiliation{\FIU}
\newcommand*{\JLAB}{Thomas Jefferson National Accelerator Facility, Newport News, Virginia 23606}
\affiliation{\JLAB}
\newcommand*{\GIESSEN}{University of Giessen, D-35392 Giessen, Germany}
\affiliation{\GIESSEN}
\newcommand*{\ANL}{Argonne National Laboratory}
\affiliation{\ANL}
\newcommand*{\ASU}{Arizona State University, Tempe, Arizona 85287-1504}
\affiliation{\ASU}
\newcommand*{\UCLA}{University of California at Los Angeles, Los Angeles, California  90095-1547}
\affiliation{\UCLA}
\newcommand*{\CSU}{California State University, Dominguez Hills, Carson, CA 90747}
\affiliation{\CSU}
\newcommand*{\CMU}{Carnegie Mellon University, Pittsburgh, Pennsylvania 15213}
\affiliation{\CMU}
\newcommand*{\CUA}{Catholic University of America, Washington, D.C. 20064}
\affiliation{\CUA}
\newcommand*{\SACLAY}{CEA-Saclay, Service de Physique Nucl\'eaire, 91191 Gif-sur-Yvette, France}
\affiliation{\SACLAY}
\newcommand*{\CNU}{Christopher Newport University, Newport News, Virginia 23606}
\affiliation{\CNU}
\newcommand*{\UCONN}{University of Connecticut, Storrs, Connecticut 06269}
\affiliation{\UCONN}
\newcommand*{\ECOSSEE}{Edinburgh University, Edinburgh EH9 3JZ, United Kingdom}
\affiliation{\ECOSSEE}
\newcommand*{\EMMY}{Emmy-Noether Foundation, Germany}
\affiliation{\EMMY}
\newcommand*{\FU}{Fairfield University, Fairfield CT 06824}
\affiliation{\FU}
\newcommand*{\FSU}{Florida State University, Tallahassee, Florida 32306}
\affiliation{\FSU}
\newcommand*{\GWU}{The George Washington University, Washington, DC 20052}
\affiliation{\GWU}
\newcommand*{\ECOSSEG}{University of Glasgow, Glasgow G12 8QQ, United Kingdom}
\affiliation{\ECOSSEG}
\newcommand*{\ISU}{Idaho State University, Pocatello, Idaho 83209}
\affiliation{\ISU}
\newcommand*{\INFNFR}{INFN, Laboratori Nazionali di Frascati, 00044 Frascati, Italy}
\affiliation{\INFNFR}
\newcommand*{\INFNGE}{INFN, Sezione di Genova, 16146 Genova, Italy}
\affiliation{\INFNGE}
\newcommand*{\ORSAY}{Institut de Physique Nucleaire ORSAY, Orsay, France}
\affiliation{\ORSAY}
\newcommand*{\BONN}{Institute f\"{u}r Strahlen und Kernphysik, Universit\"{a}t Bonn, Germany}
\affiliation{\BONN}
\newcommand*{\ITEP}{Institute of Theoretical and Experimental Physics, Moscow, 117259, Russia}
\affiliation{\ITEP}
\newcommand*{\JMU}{James Madison University, Harrisonburg, Virginia 22807}
\affiliation{\JMU}
\newcommand*{\KYUNGPOOK}{Kyungpook National University, Daegu 702-701, South Korea}
\affiliation{\KYUNGPOOK}
\newcommand*{\MIT}{Massachusetts Institute of Technology, Cambridge, Massachusetts  02139-4307}
\affiliation{\MIT}
\newcommand*{\UMASS}{University of Massachusetts, Amherst, Massachusetts  01003}
\affiliation{\UMASS}
\newcommand*{\MOSCOW}{Moscow State University, General Nuclear Physics Institute, 119899 Moscow, Russia}
\affiliation{\MOSCOW}
\newcommand*{\UNH}{University of New Hampshire, Durham, New Hampshire 03824-3568}
\affiliation{\UNH}
\newcommand*{\NSU}{Norfolk State University, Norfolk, Virginia 23504}
\affiliation{\NSU}
\newcommand*{\OHIOU}{Ohio University, Athens, Ohio  45701}
\affiliation{\OHIOU}
\newcommand*{\ODU}{Old Dominion University, Norfolk, Virginia 23529}
\affiliation{\ODU}
\newcommand*{\PITT}{University of Pittsburgh, Pittsburgh, Pennsylvania 15260}
\affiliation{\PITT}
\newcommand*{\RPI}{Rensselaer Polytechnic Institute, Troy, New York 12180-3590}
\affiliation{\RPI}
\newcommand*{\RICE}{Rice University, Houston, Texas 77005-1892}
\affiliation{\RICE}
\newcommand*{\URICH}{University of Richmond, Richmond, Virginia 23173}
\affiliation{\URICH}
\newcommand*{\UNIONC}{Union College, Schenectady, NY 12308}
\affiliation{\UNIONC}
\newcommand*{\VT}{Virginia Polytechnic Institute and State University, Blacksburg, Virginia   24061-0435}
\affiliation{\VT}
\newcommand*{\VIRGINIA}{University of Virginia, Charlottesville, Virginia 22901}
\affiliation{\VIRGINIA}
\newcommand*{\WM}{College of William and Mary, Williamsburg, Virginia 23187-8795}
\affiliation{\WM}
\newcommand*{\YEREVAN}{Yerevan Physics Institute, 375036 Yerevan, Armenia}
\affiliation{\YEREVAN}
\newcommand*{\NOWUNH}{University of New Hampshire, Durham, New Hampshire 03824-3568}
\newcommand*{\NOWUMASS}{University of Massachusetts, Amherst, Massachusetts  01003}
\newcommand*{\NOWMIT}{Massachusetts Institute of Technology, Cambridge, Massachusetts  02139-4307}
\newcommand*{\NOWECOSSEE}{Edinburgh University, Edinburgh EH9 3JZ, United Kingdom}
\newcommand*{\NOWGEISSEN}{Physikalisches Institut der Universitaet Giessen, 35392 Giessen, Germany}
\newcommand*{\NOWOHIOU}{Ohio University, Athens, Ohio  45701}
\newcommand*{\NOWSCAROLINA}{University of South Carolina, Columbia, South Carolina 29208}

\author {R.~Nasseripour} 
\affiliation{\SCAROLINA}
\author {M.H.~Wood} 
\altaffiliation[Current address:]{\NOWUMASS}
\affiliation{\SCAROLINA}
\author {C.~Djalali} 
\affiliation{\SCAROLINA}
\author {D.P.~Weygand} 
\affiliation{\JLAB}
\author {C.~Tur} 
\affiliation{\SCAROLINA}
\author {U.~Mosel} 
\affiliation{\GIESSEN}
\author {P.~Muehlich}
\affiliation{\GIESSEN}
\author {G.~Adams} 
\affiliation{\RPI}
\author {M.J.~Amaryan} 
\affiliation{\ODU} 
\author {P.~Ambrozewicz} 
\affiliation{\FIU} 
% ...
\collaboration{The CLAS Collaboration}
\noaffiliation

\date{\today}

\begin{abstract}

The photoproduction of vector mesons on various nuclei has been studied using
the CEBAF Large Acceptance Spectrometer (CLAS) at Jefferson Laboratory.
The vector mesons, $\rho$, $\omega$, and
$\phi$,  are observed via their decay to $e^+e^-$, in order to reduce the effects 
of final state interactions in the nucleus. Of particular interest are possible 
in-medium effects on the properties of the $\rho$ meson. The $\rho$ spectral 
function is extracted from the data on various nuclei, carbon, iron, and titanium, 
and compared to the spectrum from liquid deuterium, which is relatively free of 
nuclear effects.  We observe no significant mass shift for the $\rho$ meson; however, 
there is some widening of the resonance in titanium and iron, which is consistent 
with expected collisional broadening.

\end{abstract}

\pacs{11.30.Rd, 14.40.Cs, 24.85.+p}

\maketitle
\newpage

The relation between the medium modification of the masses and chiral symmetry
restoration in high density and temperature has been the subject of many studies,
both experimentally and theoretically, for about two decades. 
Spontaneous breaking of the chiral symmetry leads to a non-zero value of 
chiral condensate, $\langle q\bar{q} \rangle \sim$ 300 MeV, which is responsible for the masses of
hadrons \cite{karsh}. Indeed, most of the mass is generated dynamically. 
As the value of $\langle q\bar{q} \rangle$ is predicted to decrease with increasing
temperature and density, in-medium modification of the masses are expected in hot/dense
mediums \cite{gellman,kunihiro,weise,meissner,klimt,shyryak,brown,hatsuda,herrmann,rapp}. 
Quantitative understanding of the dynamics in the non-perturbative regime of
QCD is rather incomplete and therefore experimental data is essential.
If these modifications exist, they could also be an indication of a transition from hadronic 
matter to a deconfined quark-gluon plasma \cite{qgp}.  

To investigate these modifications, high temperature and density can be achieved by
experiments using heavy-ion colliders. The first indication of a possible medium modification of the
$\rho$ meson came from the CERES~\cite{ceres1} and HELIOS/3~\cite{helios} collaborations at CERN in 1995.  
CERES reported on the measurement of low mass $e^+ e^-$ pairs from p-Au and Pb-Au collisions.
While their proton-induced data could satisfactorily be accounted for by summing 
various hadron decay contributions, an enhancement over the hadronic contributions 
was observed for the Pb-Au data in the mass range between 300 and 700 MeV. 
However, the resolution and statistics of the data did not allow a quantitative
measure of a mass shift or a change in the width of the $\rho$ vector meson.
Recently, following an upgraded experiment, CERES reported a broadening in the mass of 
the $\rho$ vector meson due to the baryonic interaction in the low mass region \cite{ceres2}.
The result of di-muon measurement in In-In collision in NA60 experiment at CERN 
has also shown a considerable broadening (doubling of the width) of the $\rho$ spectrum, 
while no shift in the mass was observed \cite{na60,na60-1,na60-2}. 

Although a large medium effect is expected in the heavy-ion reactions, the results
are integrated over a wide range of temperature and density, many channels contribute,
and the reactions proceed far from equilibrium. 
This makes an exact interpretation of results difficult, and 
the connection between chiral symmetry restoration and in-medium modifications remains unclear 
\cite{hades,phenix,ceres3}.
However, the in-medium effects for the vector mesons at normal nuclear density and
zero temperature are predicted to be so large that they can be studied by 
fixed-target experiments involving elementary reactions \cite{mosel}.

Brown and Rho \cite{brown} 
start from an effective Lagrangian at low energy and zero density. 
They apply the same Lagrangian at nuclear density but with masses and coupling 
constants that are modified according to the symmetry constraints of QCD 
(e.g., chiral symmetry). This model predicts an in-medium scaling law that results in 
a decrease in the mass of vector mesons by $20\%$.

Hatsuda and Lee \cite{hatsuda}, based on QCD sum rule calculations,
obtain mass changes of the vector mesons in the nuclear medium.
Their calculations result in a linear decrease of the masses as a function of 
density introducing a mass shift parameter $\alpha$:
\begin{equation}
{m_{VM}(\rho) \over m_{VM}(\rho=0)} = 1 - \alpha { \rho \over \rho_0}, \;\;\;\;\; \alpha=0.16\pm0.06,
\label{eq:alpha}
\end{equation} 
where $m_{VM}$ is the mass of the vector meson, 
$\rho_0$ indicates nominal nuclear density (0.16 fm$^{-3}$), and $\rho=0$ indicates the vacuum.

Models based on nuclear many-body effects predict a broadening in the width of
the $\rho$ meson with increasing density. This prediction is based on the assumption
that many-body excitations may be present with the same quantum numbers and
can be mixed with the hadronic states \cite{herrmann,rapp,mosel2,wambach,oset,madeline}. 
Furthermore, due to the uncertainty of the coupling constants as a function of 
density, the branching ratios are expected to change in the nuclear medium and 
also distort the invariant mass spectrum of the resonance \cite{eichstaedt}. 

An observation of a medium-modified vector meson invariant mass decrease 
($\alpha$ =  0.09$\pm$0.002) has been 
claimed by a KEK-PS collaboration in an experiment where 12 GeV protons were incident 
on nuclear targets, and the $e^+e^-$ pairs were detected~\cite{kek,kek2,kek-new}.  
Very recently, the Crystal Barrel/TAPS collaboration has reported a 14$\%$ downward shift 
in the mass of the $\omega$, where the analysis focused on the $\pi^0\gamma$ decay 
of low-momentum $\omega$ mesons photoproduced on a nuclear target~\cite{taps}.

The data for the present study were taken in 2002 using the CEBAF accelerator and the 
CLAS detector located in Hall-B of Jefferson Laboratory. A comprehensive 
description of CEBAF, the Continuous Electron Beam Accelerator Facility, can be 
found in the Ref.~\cite{lee01}, while Ref.~\cite{mec03} gives a
detailed account of CLAS, the CEBAF Large Acceptance Spectrometer, and 
other Hall-B equipment. The incident bremsstrahlung photon beam on the  
target was produced from a primary electron beam of 3 GeV in energy for 
the first 2/3 of the run time, and a primary electron beam of 4 GeV for 
the last 1/3 of the run time.

The \v{C}erenkov counters (CC) in combination with the electromagnetic 
calorimeters (EC) were the two most crucial CLAS components for this experiment. 
The EC and CC cover the forward part of the CLAS detector, subtending 
scattering angles of $8^{\circ} < {\theta} < 45^{\circ}$.
The $e^+e^-$ event selection and the rejection of the very large $\pi^+\pi^-$ 
background were done through cuts on the EC and the CC. The pion rejection (or 
misidentification) factor is determined to be of the order of 10$^{-7}$ for
a two-track measurement \cite{longpaper}. 

The target contained a liquid-deuterium (LD$_2$) cell and six solid foils, 
each with a 1.2 cm diameter.  Four of the foils were carbon, one was titanium, and one
iron.  The total thickness of the deuterium, and the four carbon targets was each 
$1$ g/cm$^2$, while the titanium and iron targets were each ${1 \over 2}$ g/cm$^2$. 
The atomic weights of iron and titanium were close enough that the 
data from these two targets were combined to increase the statistics.
The separation between targets was $2.5$ cm; the target nucleus was determined by the 
reconstructed position of the production vertex; and the CLAS vertex reconstruction
resolution was 0.3 cm.

The object of this study is the invariant mass of $e^+ e^-$ from the decay 
of the vector mesons $\rho$, $\omega$, and $\phi$.  
This branching ratio for vector mesons 
into $e^+ e^-$ is of the order of $10^{-4}-10^{-5}$.  Other physical 
processes also contribute to the background, for example 
$\omega \rightarrow \pi^0 e^+ e^-$, $\eta \rightarrow \gamma e^+ e^-$, 
and  $\pi^0 \rightarrow \gamma e^+ e^-$ (Dalitz decay). In the case of 
the $\eta$ and $\pi^0$, the $e^+ e^-$ mass is below the region of interest.
In the case of the $\omega$, the Dalitz decay is also included in the expected 
spectrum, with this mode tied to the $e^+ e^-$ mode.
The background from $\gamma A \rightarrow \pi^0 \pi^0 X$ 
with both pions decaying via the Dalitz mode is also considered.
In this case, the $e^+$ may be detected from one pion and the $e^-$ detected 
from the other.  This process was simulated with the known $\pi^0 \pi^0$ cross 
section \cite{ulrike} and angular distributions, and its contribution determined
to be small (0.02$\%$).  In addition, Bethe-Heitler  $e^+ e^-$ pairs have been simulated with 
the expected cross section and mass distribution, and also found to be negligible ($<$0.01$\%$).

\begin{figure}[htbp]
\vspace{3.4cm}
\includegraphics{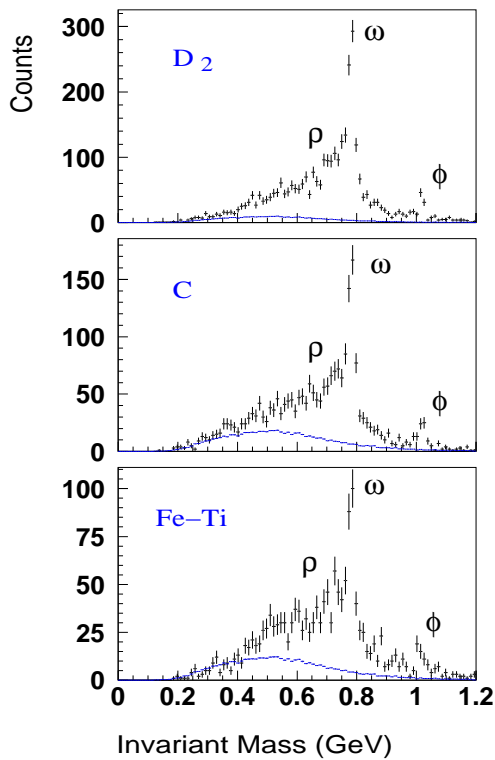}
\vspace{8.7cm}
\caption[Estimated combinatorial background.]
{(Color online) Normalized combinatorial background (blue) for all 
and individual targets compared to data (black). }
\label{fig:comb-tg}
\end{figure}

Lepton pair production also has a background of random combinations of $e^+ e^-$ pairs 
due to uncorrelated sources occurring within the same 2 ns CEBAF 
beam bucket. The most salient feature of the uncorrelated sources is that they 
produce the same-charge lepton pairs as well as oppositely charged pairs.
The same-charge pairs ($e^+e^+$ and $e^-e^-$) 
provide a natural normalization of the uncorrelated background.
This is also true for the measurement of opposite-sign pions or muons for which 
the combinatorial method has also been used in the past \cite{pions,muons}.
This method has also been used in the extraction of resonance signals \cite{res} and 
proton femtoscopy of $eA$ interactions \cite{stepan}.

The combinatorial background is determined by an event-mixing
technique. The electrons of a given event are combined with positrons of another
event, as the two samples of electrons and positrons are completely uncorrelated. 
This produces the phase-space distribution where electrons and positrons are 
actually from different processes but lying in the same event. 
  
\begin{figure}[ht]
\vspace{2.7cm}
\includegraphics{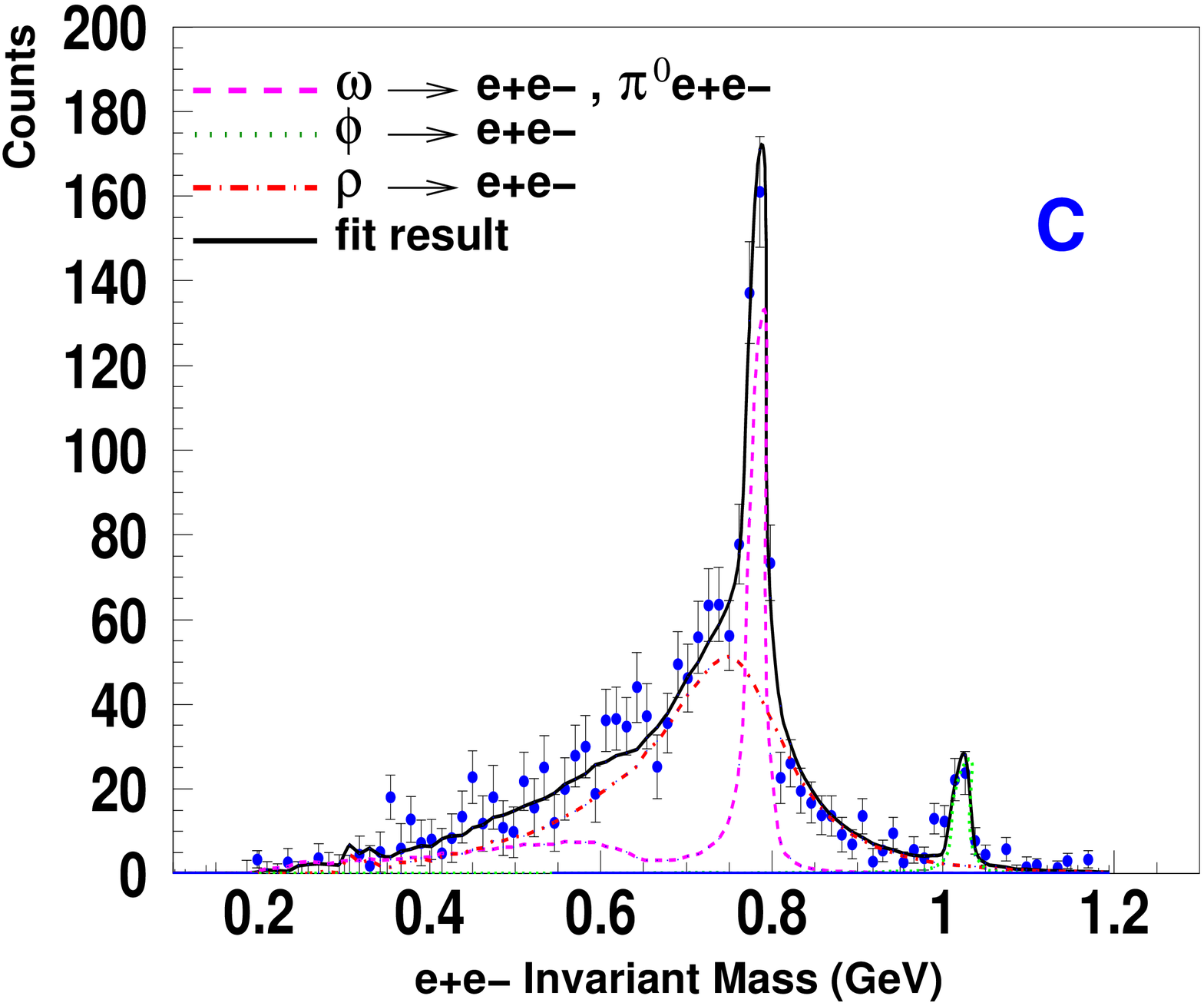}
\vskip 6.2cm
\includegraphics{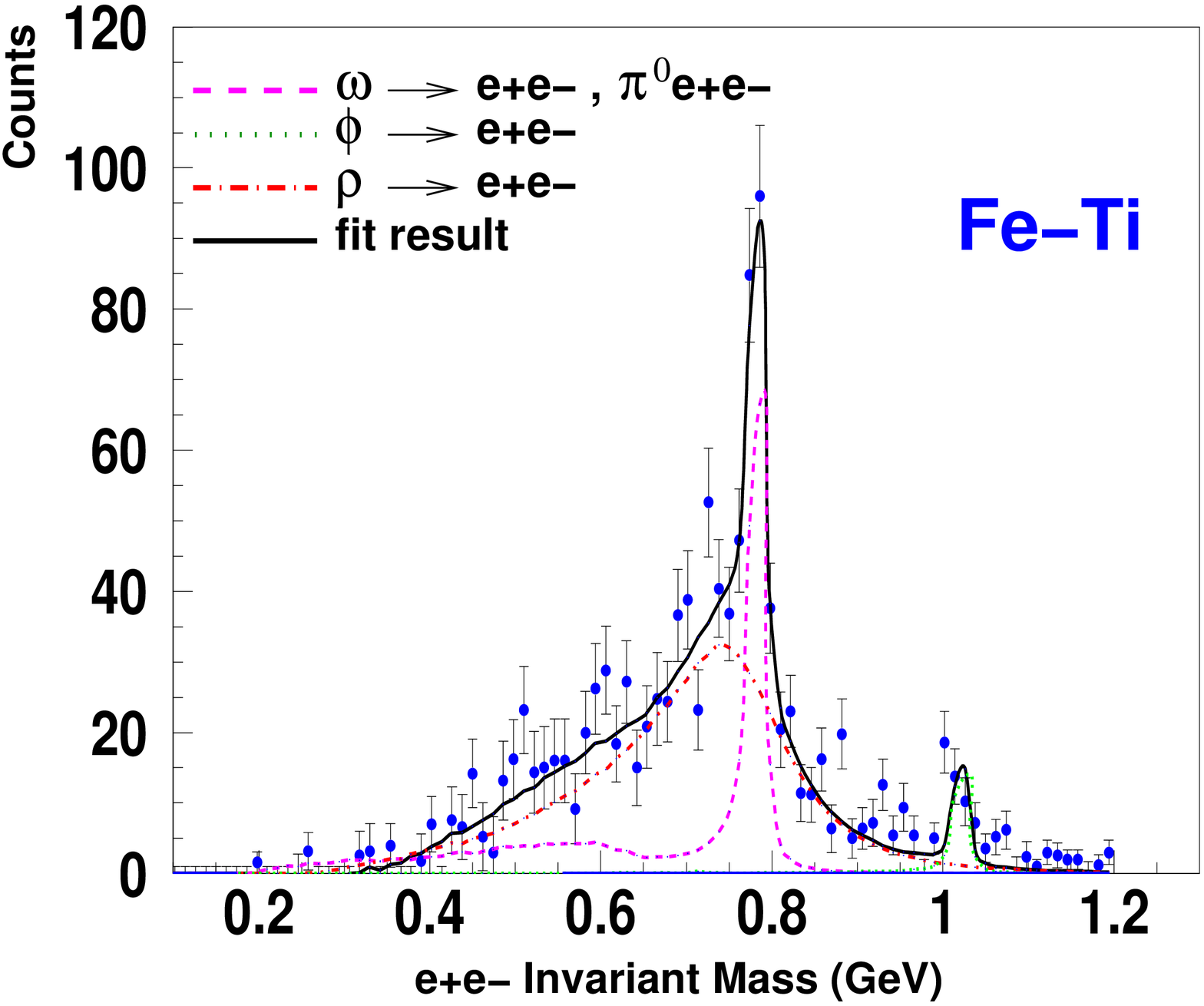}
\vspace{4.7cm}
\caption[Carbon and Fe-Ti targets.]{(Color online) Result of the fit to the $e^+e^-$ 
invariant mass obtained for the carbon (top) and Fe-Ti (bottom) data. The curves are  
calculations by the BUU model \cite{effenberger1,effenberger2} for various $e^+ e^-$ channels.}
\label{fig:resultfe}
\end{figure}
\begin{figure}[t]
\vspace{0.8cm}
\includegraphics{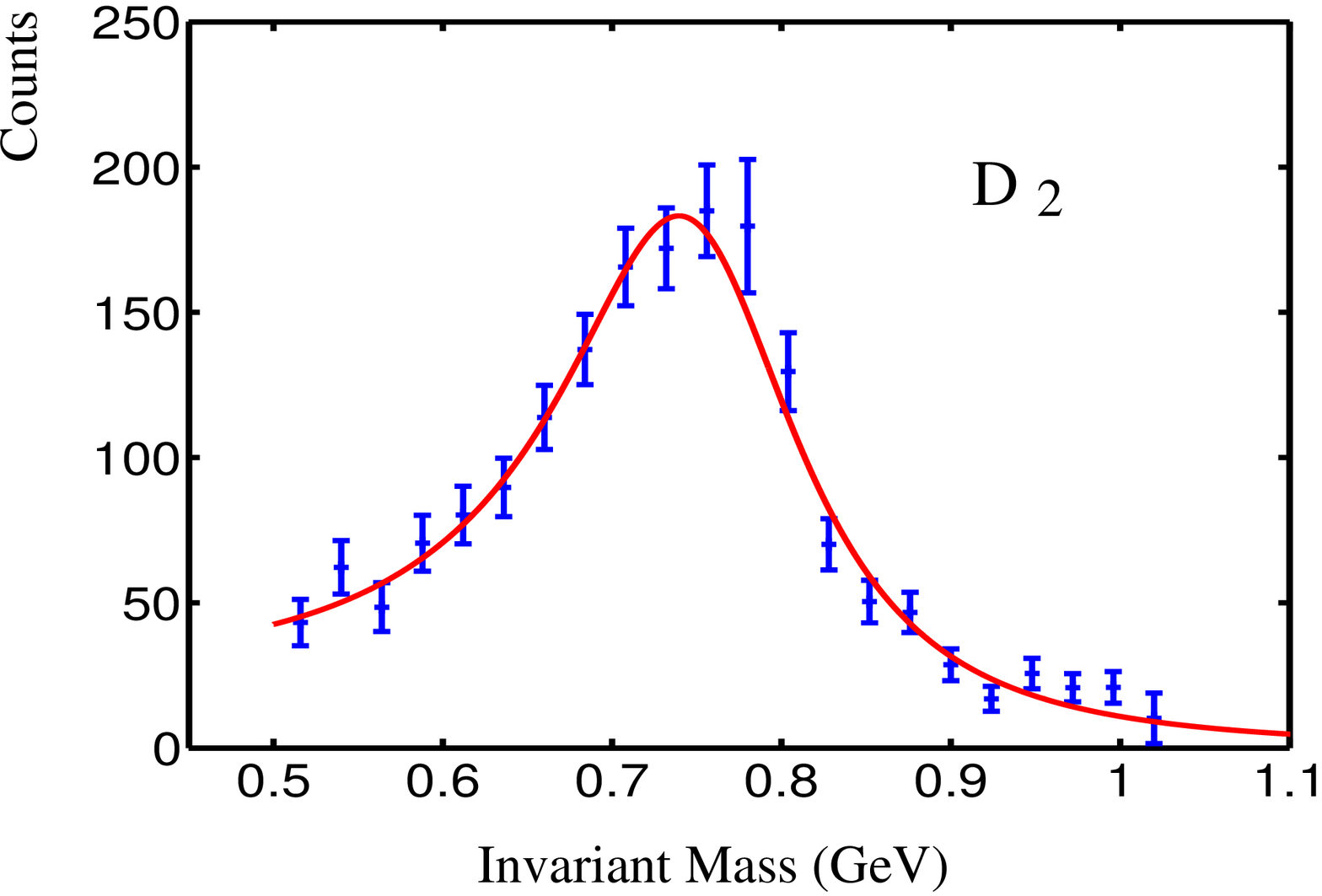}
\vskip 4.7cm
\includegraphics{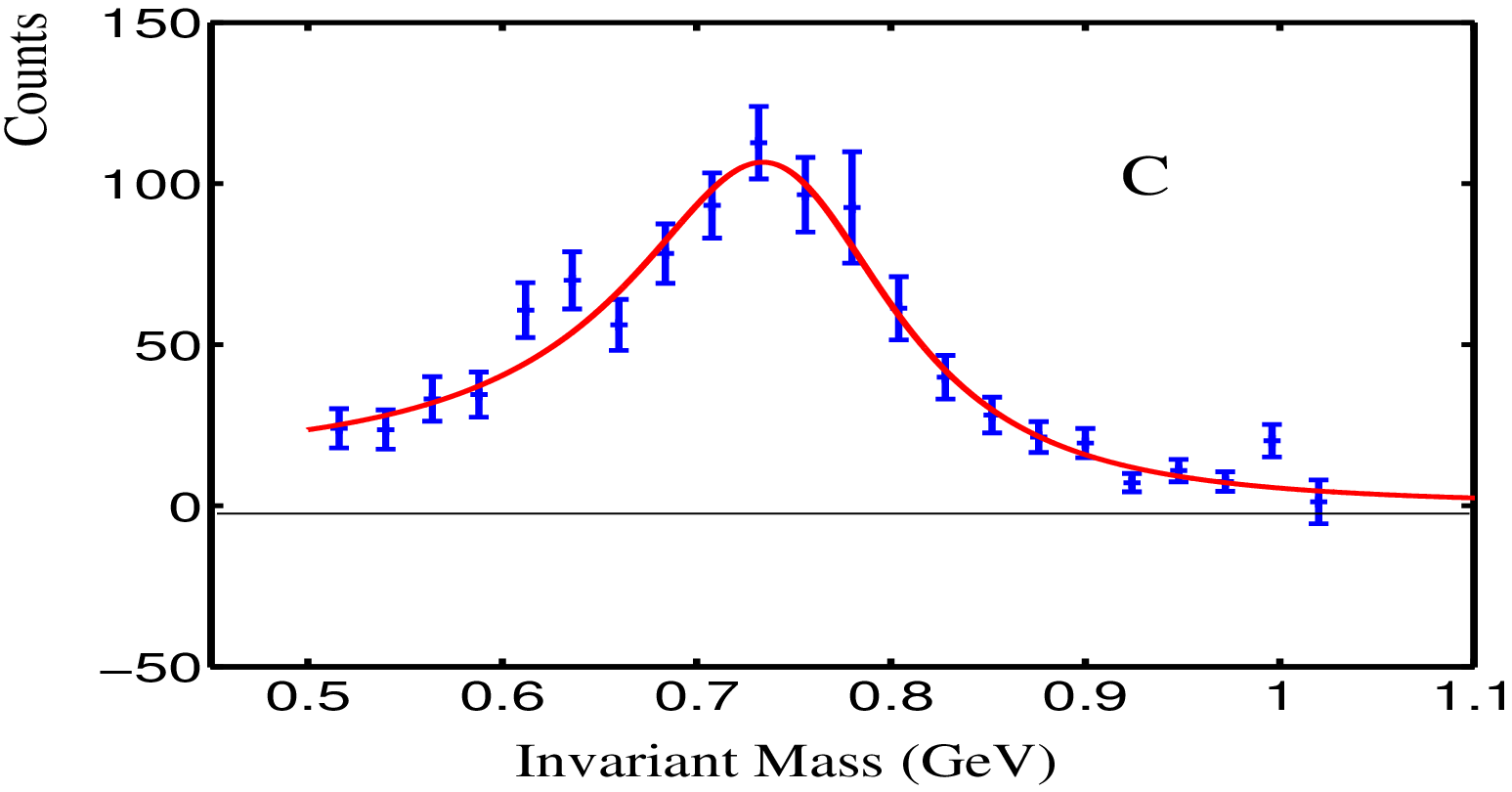}
\vspace{4.7cm}
\includegraphics{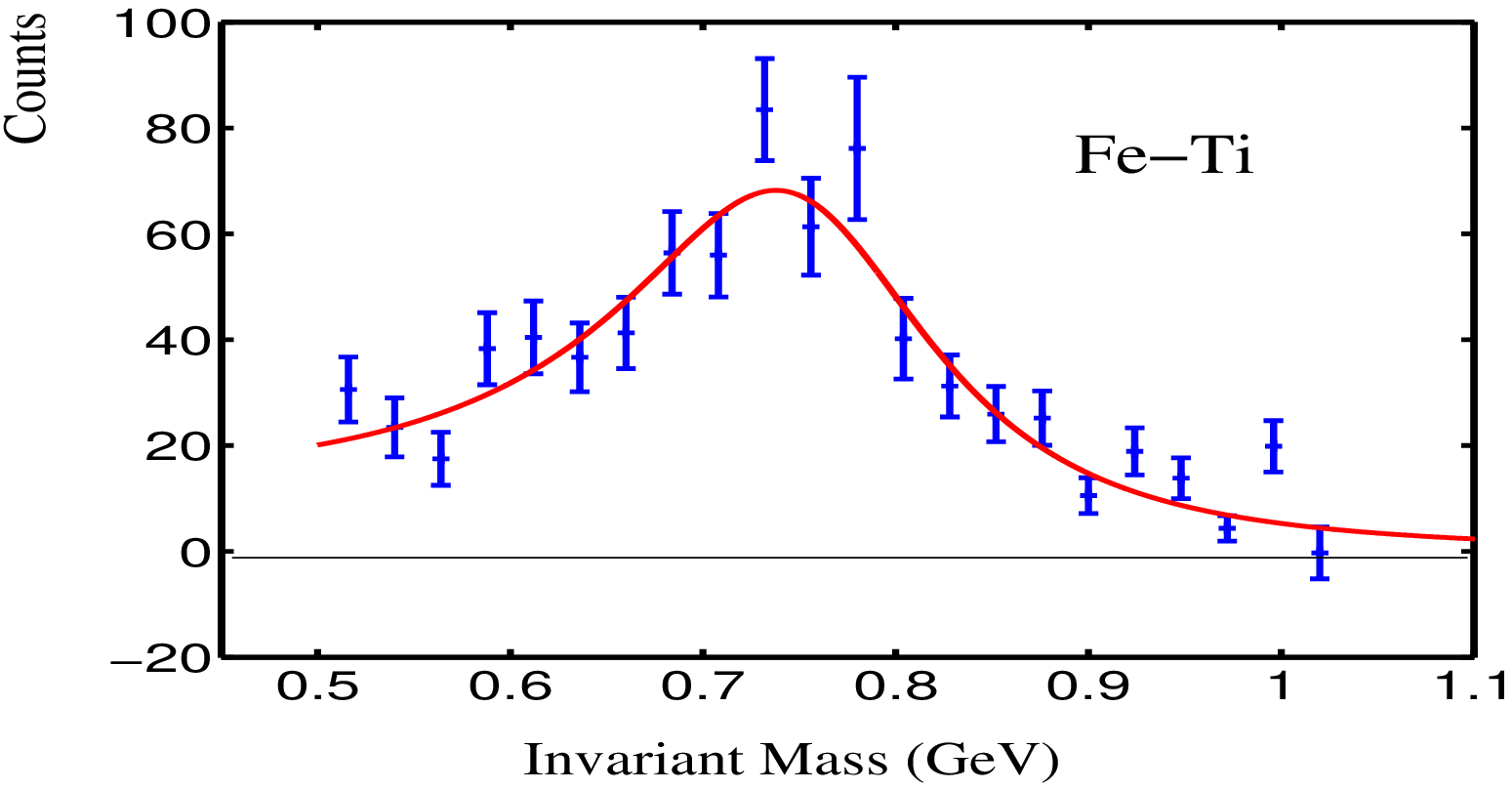}
\vspace{4.7cm}
\caption[Simultaneous fits to $\rho$ mass spectra, D$_2$,carbon,and iron/titanium.]
{The result of a simultaneous fit to the $\rho$ mass spectra and the ratios for D$_2$ (top), C (middle), 
and Fe-Ti (bottom).} 
\label{fig:fe-fit}
\end{figure}
\begin{figure}[!]
\vspace{1.0cm}
\includegraphics{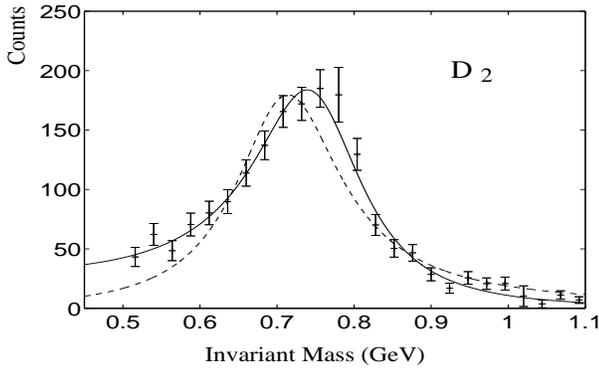}
\vspace{4.7cm}
\caption[]
{Fits to the $\rho$ mass distribution from D$_2$ data using a Breit-Wigner 
function (dashed line) and Breit-Wigner/$\mu^3$ (solid line).} 
\label{fig:BWfit}
\end{figure}

The mixed opposite-charged leptons chosen from samples
of uncorrelated events were used to estimate the shape of the combinatorial background. 
This distribution was then normalized to the number of expected opposite-charge pairs.
The result is shown in Fig. \ref{fig:comb-tg} for the individual targets.

To simulate each physics process, a realistic model was employed and corrected for 
the CLAS acceptance. The events were generated using a code based on a semi-classical 
Boltzmann-Uehling-Uhlenbeck (BUU) transport model developed by the Giessen group 
that treats photon-nucleus reactions as a two-step 
process~\cite{mue02}. In the first step, 
the incoming photons react with a single nucleon, taking into account various nuclear effects,  
e.g. shadowing,
Fermi motion, collisional broadening, Pauli blocking, and nuclear binding. 
Then in the second step, the produced particles are propagated explicitly through the nucleus 
allowing for final-state interactions, governed by 
the semi-classical BUU transport equations. A rather complete 
treatment of the $e^+e^-$ pair production from $\gamma A$ reactions at Jefferson Lab energies 
using this code can be found in Ref.~\cite{bra99}.

\renewcommand{\baselinestretch}{0.2}
\begin{table}[h]
\begin{center}
\begin{tabular} {|c|c|c|c|c|} \hline
Target  	& Mass 			& Width  		& Mass  	& Width 	\\ 
        	&  CLAS data    		&   CLAS data   		&  BUU       	& BUU        	\\ \hline \hline
D$_2$		& 770.3$\pm$3.2 	& 185.2$\pm$ 8.6   	& 774.5$\pm$4.9	& 160.1$\pm$10.2	\\ \hline
C       	& 762.5$\pm$3.7 	& 176.4$\pm$ 9.5        & 773.8$\pm$0.9	& 177.6$\pm$\thinspace\thinspace2.1	\\ \hline
Fe-Ti    	& 779.0$\pm$5.7 	& 217.7$\pm$14.5      & 773.8$\pm$5.4	& 202.5$\pm$11.6	\\ \hline
\hline
\end{tabular}
\renewcommand{\baselinestretch}{1}
\vspace{0.2cm}
\caption[The mass and width  of the $\rho$ meson obtained from the
simultaneous fits to the mass spectra.]{The mass and width (in MeV) of the $\rho$ meson obtained from the
simultaneous fits to the mass spectra for each target and the ratio to D$_2$ compared
to the result of the BUU simulations.
The masses are consistent with the PDG values (770.0 $\pm$ 0.8 MeV) and the widths are consistent
with the collisional broadening (150.7 $\pm$ 1.1 MeV).}
\renewcommand{\baselinestretch}{1}
\label{tab:c-fe-fit}
\end{center}
\end{table}
\renewcommand{\baselinestretch}{1}
The expected combinatorial background distributions are subtracted from the 
$e^+ e^-$ effective mass distributions.  
The peaks of the $\omega$ and $\phi$ vector mesons are prominent  
in the invariant mass spectra, and one can determine the normalization of these narrow peaks 
rather easily. 
The shape of these peaks and the $\omega$ Dalitz channel 
are well described by the BUU model where the ratio of $\omega$ to  $\omega$ Dalitz
decay was also fixed to their branching ratios. These distributions were fit to the data, 
then the resulting normalized heights were subtracted, leaving just the experimental 
spectra of the $\rho$ mass.  
These fits for carbon and iron/titanium are shown in Fig.~\ref{fig:resultfe}.

The extracted $\rho$ mass distributions 
are then fit with the exact spectral functional form 
obtained from calculating the cross section of production of the $\rho$ meson
including the leptonic decay width ~\cite{bra99,guo,oconnell}.
The results of the fits superimposed on the data are shown in Fig.~\ref{fig:fe-fit}, 
and the results are tabulated in Table~\ref{tab:c-fe-fit}. 
The fits describe the data very well, and while the width of the $\rho$ meson 
is consistent with the natural width of 150 MeV and collisional broadening \cite{bugg}--
that is also included in the BUU calculations-- 
it is not compatible with the doubling of the $\rho$ width 
reported by NA60 \cite{na60}. 

A good approximation for the $\rho$ spectral function used to fit the data is a Breit-Wigner 
function 
divided by $\mu^3$, where $\mu$ is the mass of the $e^+e^-$ pair. The factor $1/\mu^3$ comes from the 
photon propagator ($1/\mu^4$) that couples to the $\rho$ meson in the dilepton decay diagram,
multiplied by a phase-space factor. 
Indeed the fits to the Breit-Wigner/$\mu^3$ rather than a simple Breit-Wigner function 
describe the data very well. For example, for the D$_2$ target a simple Breit-Wigner fit 
gives a $\chi^2$ per degrees of freedom = 3.9 while a Breit-Wigner/$\mu^3$
gives a  $\chi^2$ per degrees of freedom = 1.08 (see Fig. \ref{fig:BWfit}). The sensitivity of the 
fits to the 1/$\mu^3$ factor indicates that the systematic uncertainties in the 
background subtraction are insignificant, and the $\rho$ spectra
are cleanly extracted.  
Similar results are obtained for the heavier targets, C and Fe-Ti, where the uncorrelated 
background is proportionally larger. 

Based on the notation of Ref.~\cite{hatsuda} and Eq. \ref{eq:alpha}, we obtain 
the shift parameter 
$\alpha$ =  0.02$\pm$0.02 for the Fe-Ti target with $\rho$ momenta ranging from
0.8 to 3.0 GeV. This is consistent 
with no significant mass shift predicted by the calculations  
of Ref.~\cite{wambach,oset} and those of Ref.~\cite{mue02,mue04} at $\rho$ vector 
meson momenta $>$ 1 GeV.

 The total systematic uncertainty for the measured $\alpha$ due to various
sources is estimated to be $\Delta \alpha$ = $\pm$0.01 \cite{longpaper}.

Our result sets an upper limit of $\alpha$ = 0.04 with a 95$\%$ confidence 
level. This does not favor the prediction of Refs.~\cite{brown} and \cite{hatsuda} for
a 20$\%$ mass shift and $\alpha$ =  0.16$\pm$0.06 respectively, and 
is significantly different from other similar experiments~\cite{kek,kek2,kek-new},
where $\alpha$ =  0.092$\pm$0.002, 
with no broadening in the width of the $\rho$ meson. Our results are also not 
necessarily inconsistent
with the result of the experiment in Ref. \cite{taps} that measures a $-$14$\%$ 
shift in the mass of the $\omega$ meson, 
since different medium modification mechanisms are indeed
expected for $\rho$ and $\omega$ mesons \cite{omega1,omega2}.

The extracted experimental $\rho$ mass spectrum with the unique characteristic of 
electromagnetic interactions in both the production and decay, 
is well described by the $\rho$ functional form obtained from the exact calculations 
given in Refs.~\cite{guo,bra99,oconnell} with no modification
beyond standard nuclear many-body effects. 
With the availability of more sophisticated theoretical models and improved analysis 
techniques, future experiments with higher statistics
are expected to make a conclusive statement about the momentum dependence of the 
in-medium modifications and the nature of the QCD vacuum.


We would like to thank the staff of the Accelerator and Physics Divisions at Jefferson
Laboratory who made this experiment possible. This work was supported in part by 
the U.S. Department of Energy and National Science Foundation, the Research Corporation,
the Italian Instituto Nazionale de Fisica Nucleare, 
the French Centre National de la Recherche Scientifique and Commissariat \'a l'Energie Atomique, 
the Korea Research Foundation, the U.K. Engineering and Physical Science Research Council. 
Jefferson Science Associates (JSA) operates the Thomas Jefferson National Accelerator 
Facility for the United States Department of Energy under contract DE-AC05-06OR23177.

\end{document}